\documentclass[aps,fleqn,preprint]{revtex4-1}

\usepackage{amsmath}
\usepackage{graphicx}
\usepackage{epsfig}
\usepackage{subfigure}
\usepackage{amsfonts}
\usepackage{amssymb}
\usepackage{amsthm}
\usepackage{newlfont}
\usepackage{epstopdf}

\usepackage[utf8]{inputenc}

\usepackage{dcolumn}% Align table columns on decimal point
\usepackage{bm}% bold math

\usepackage[section] {placeins}

\begin{document}

\title{An excess chemical potential for binary hard-sphere mixtures from integral equation theory }

\author{Banzragch Tsednee}
\author{Tsogbayar Tsednee}
\author{Tsookhuu Khinayat}
\affiliation{Institute of Physics and Technology, Mongolian Academy of Sciences, Peace Ave 54B, 13330 Ulaanbaatar, Mongolia }

\begin{abstract}

We solve the site-site Ornstein-Zernike equation using the Percus-Yevick closure for binary hard-sphere mixture. We calculate an excess chemical potential for the mixture's diameter ratios of 0.3, 0.5, 0.6 and 0.9, and at packing fraction of 0.49 using the analytical expression. Our numerical results are in good agreement with those in the literature.

%This is the first time the Floquet technique combined with the complex absorbing potential has been employed for photoionization cross sections of the ion. We also report on ionization rates of the ion for different internuclear distances with intensity of $I=5\times 10^{13}\, W/cm^{2}$. We compare our findings of from calculations carried out in both gauges with those of previous calculations.   

\end{abstract}

\pacs{Valid PACS appear here}% PACS, the Physics and Astronomy

\maketitle

\section{Introduction}

In statistical thermodynamics a direct calculation of chemical potential is a difficult task than computing other quantities, such as an internal energy, the pressure, and the correlation functions determining the structures of the physical systems. Various theoretical methods have been extensively employed to calculate the chemical potential, such as, the Widom's test particle insertion method~\cite{Widom63} or its modifications~\cite{Adams74, Attard93} and the scaled-particle Monte-Carlo (SP-MC) method~\cite{Barosova96, Borowko00, Martin18}. 

In theoretical studies for liquid an integral equation (IE) theory has been widely used due to its low computational cost and a fact that the theory is a thermodynamically complete, when combined with an appropriate closure equation~\cite{Ornstein14, Percus58, Chandler72, Hirata81, Perkyns92, Kovalenko99, Luchko16, Lebowitz58, Tsogbayar19}. From the IE approach, calculating the chemical potential is almost straightforward employing an analytical expression based on the correlation functions.  Therefore, in this work our purpose is to calculate the chemical potential for binary hard-sphere mixtures using the IE approach with the Percus-Yevick closure [8]. We will also compare our numerical findings with the SP-MC results obtained by Baro\v{s}ov\'{a} et. al., \cite{Barosova96}. 

The paper consists of four sections. The first section presents the Introduction, while the second and the third sections discuss a brief description of theoretical formulation and numerical results, respectively. Then a conclusion follows.

\section{Integral equation theory}

For the atomic mixture system, the site-site Ornstein-Zernike equation (SSOZ) \cite{Chandler72, Hirata81, Perkyns92, Kovalenko99} which establishes a relation between the total correlation function $h_{ij}$  and the direct correlation function $c_{ij}$ in the $\mathbf{k}$-space can be written in the form
\begin{equation}\label{ssoz}
 \tilde{h}_{ij} = \tilde{c}_{ij} + \sum_{i'} \rho_{i'} \tilde{c}_{i i'} \tilde{h}_{i' j}, \quad (i, i', j = 1, 2), 
\end{equation}
where $\rho_{i} = x_{i} \rho$ is a number density, and $x_{i}$ is the mole fraction of component $i$, and $\rho = \sum_{i} \rho_{i}$ is the total number density. The tildes in Eq.~(\ref{ssoz}) denote a Fourier transformation. Since we do not know both correlation functions in Eq.~(\ref{ssoz}), we cannot solve this equation directly. Therefore, to solve this equation, we need an auxiliary equation which we call a closure relation, and then they can be solved self-consistently. A general closure for the mixture may be given in the form 
\begin{equation}\label{hsu}
 h_{ij} = \exp [-\beta u_{ij} + h_{ij} - c_{ij} + B_{ij}] - 1. 
\end{equation}
Here $u_{ij}$ is an interaction potential, which in this work is given by
\begin{equation}\label{u_ij}
	u_{ij} (r) = 
   \begin{cases}
            r  <   \sigma_{ij} \\
            r \geq \sigma_{ij}, 
   \end{cases}
\end{equation}
where $\sigma_{ij} = \frac{1}{2}(\sigma_{i} + \sigma_{j})$, and $\sigma_{i}$ is a diameter of component $i$, and $\beta = 1/k_{\mathrm{B}} T$, $k_{\mathrm{B}}$ is a Bolztmann's constant and $T$ is the temperature for the system. .

The Percus-Yevick (PY) bridge function~\cite{Percus58} which we will use in our calculation has a form
\begin{equation}\label{py_br}
 B_{ij} = \ln (1+\gamma_{ij}) - \gamma_{ij},
\end{equation}
where $\gamma_{ij} = h_{ij} - c_{ij}$ is the indirect correlation function.  We note that in this PY approximation a solution for an IE for the hard-sphere systems can be found in an analytical form \cite{Lebowitz58}.
Once we have the correlation functions describing the structure of the mixture, we can compute an excess chemical potential for component $i$ using a following expression
\begin{equation}\label{mu_ex}
 \beta \mu^{ex}_{i} = \sum_{j} \rho_{j} \int \Big[\Big(\frac{1}{2}h^{2}_{ij} - c_{ij} - \frac{1}{2}h_{ij}c_{ij} \Big)  + \Big(1 + \frac{2}{3} h_{ij} \Big) B_{ij}\Big] d\mathbf{r}.
\end{equation}
Note that a derivation of this expression can be performed similarly as it was obtained for the pure Lennard-Jones fluid by Tsogbayar and Luchko~\cite{Tsogbayar19}. Moreover, the expression can be used for any bridge functions since it does not require the explicit forms of them.

\section{Results and Discussion}

In our calculation we chose a component 1 as a reference particle which is a larger component of the mixture, that is $(\sigma_{1} > \sigma_{2})$, and a packing fraction $\eta$ is given as
\begin{equation*}\label{eta_mf}
 \eta = \frac{\pi}{6} \rho (x_{1} \sigma^{3}_{1} + x_{2} \sigma^{3}_{2}),
\end{equation*}
where $x_{2} = 1 - x_{1}$  is the mole fraction for component 2. We solve an Eq. (\ref{ssoz}) numerically using the Picard iteration method. We performed numerical calculations for four different values of diameter ratio $\sigma_{2}/\sigma_{1}: 0.3, 0.5, 0.6, 0.9$ and at $\eta = 0.49$. This value of the packing fraction indicates that our calculation can be done in the stable liquid region since the value is slightly lower that the freezing packing fraction for pure hard spheres \cite{Barosova96}. 
We do our calculation as function of the mole fraction $x_{1}$ of the larger component. In all calculations a number of grid points is $N = 2^{15}$, and a length interval is $ L = 32 \sigma_{1}$. The convergence criterion for a successive two iterations of the indirect correlation function is set to $10^{-6}\sigma_{1}$.

The columns 2 and 3 in Table I show our numerical results of an excess chemical potential $\beta \mu^{ex}_{1}$, for various values of the mole fraction $x_{1}$ and at $\eta = 0.49$. We placed results of the SP-MC method~\cite{Barosova96} in the columns 3 and 5. When $\sigma_{2}/\sigma_{1} = 0.3$, and for the smaller values of $x_{1}$ our calculated values show small discrepancies, however they reduce with an increase of the mole fraction of the larger component. For $\sigma_{2}/\sigma_{1}=0.9$, our values placed in column 4 are close to the SP-MC data shown in the last column. We note that the number of particles used in the conventional canonical ensemble Monte-Carlo method of Ref.~\cite{Barosova96} is $1728$, however, since our IE method is an implicit method, it is independent on number of particles involved in the system, that is, we do not care about number of particles in the IE formulation.
\begin{table}[h]
\caption{ A comparison of our calculated values for the excess chemical potential $\beta \mu^{ex}_{1}$ with those obtained by the SP-MC method~\cite{Barosova96} for $\sigma_{2}/\sigma_{1} = 0.3$, and $0.9$ and at $\eta = 0.49$.}
\begin{center}
{\scriptsize
\begin{tabular}{l@{\hspace{2mm}}c@{\hspace{2mm}}c@{\hspace{2mm}}c@{\hspace{2mm}}c@{\hspace{2mm}}c@{\hspace{2mm}}c@{\hspace{2mm}}c@{\hspace{2mm}}c@{\hspace{2mm}}c@{\hspace{2mm}}c@{\hspace{2mm}}c@{\hspace{2mm}}c@{\hspace{2mm}}c@{\hspace{2mm}} }
\hline\hline
& \multicolumn{1}{c}{$\sigma_{2}/\sigma_{1} = 0.3$ }  &  & \multicolumn{4}{c}{$\sigma_{2}/\sigma_{1} = 0.9$ }  \\
%\cline{2-2}
 $x_{1}$ & $\beta \mu^{ex}_{1}$ (Eq.~\ref{mu_ex}) & Ref.~\cite{Barosova96} &  & & $\beta \mu^{ex}_{2}$ (Eq.~\ref{mu_ex}) & Ref.~\cite{Barosova96}  \\
\hline
$0.0625$ & $63.44$ & $66.75$ & & & 19.14  \\
0.125 & 39.29 &	41.21 & & & 18.83 \\
0.25	& 26.02 &	27.03 & & & 18.23 & 18.71\\
0.5	& 19.17 &	19.67 & & & 17.22 & 17.61\\
0.75	& 16.89 &	17.20 & & & 16.40 & 16.70 \\
0.95	& 15.88 & & & & 15.83	\\
\hline\hline
\end{tabular} }
\end{center}
\end{table}
In Table II we present a comparison of our calculated values for $\beta \mu^{ex}_{2}$ with available values obtained by the SP-MC method \cite{Barosova96} for $\sigma_{2}/\sigma_{1} = 0.3$  and $0.9$, and at $\eta =  0.49$. All our findings in this table are close to an accurate SP-MC values \cite{Barosova96}.
\begin{table}[h]
\caption{ The same as in Table I, but for an excess chemical potential $\beta \mu^{ex}_{2}$.}
\begin{center}
{\scriptsize
\begin{tabular}{l@{\hspace{2mm}}c@{\hspace{2mm}}c@{\hspace{2mm}}c@{\hspace{2mm}}c@{\hspace{2mm}}c@{\hspace{2mm}}c@{\hspace{2mm}}c@{\hspace{2mm}}c@{\hspace{2mm}}c@{\hspace{2mm}}c@{\hspace{2mm}}c@{\hspace{2mm}}c@{\hspace{2mm}}c@{\hspace{2mm}} }
\hline\hline
& \multicolumn{1}{c}{$\sigma_{2}/\sigma_{1} = 0.3$ }  &  & \multicolumn{4}{c}{$\sigma_{2}/\sigma_{1} = 0.9$ }  \\
%\cline{2-2}
 $x_{1}$ & $\beta \mu^{ex}_{1}$ (Eq.~\ref{mu_ex}) & Ref.~\cite{Barosova96} &  & & $\beta \mu^{ex}_{2}$ (Eq.~\ref{mu_ex}) & Ref.~\cite{Barosova96}  \\
\hline
$0.0625$ & $5.10$ & $5.146$ & & & 15.34  \\
0.125 & 3.76 &	3.766 & & & 15.08 \\
0.25	& 2.97 &	2.957 & & & 14.62 & 14.97 \\
0.5	& 2.55 &	2.519 & & & 13.82 & 14.11\\
0.75	& 2.40 &	2.368 & & & 13.17 & 13.42 \\
0.95	& 2.33 & & & & 12.73	\\
\hline\hline
\end{tabular} }
\end{center}
\end{table}
Once we compute the excess chemical potential, we may obtain directly the logarithm of the activity coefficient using a following equation~\cite{Barosova96}:
\begin{equation}\label{logact}
 \ln \gamma_{i} = \beta \mu^{ex}_{i}(x_{i}, \eta) - \beta \mu^{ex}_{i}(1, \eta).
\end{equation}
In Fig. 1 we show a logarithm of the activity coefficient calculated with an Eq. 6 for the four different diameter ratios as function of $x_{1}$ at constant $\eta = 0.49$ for the smaller component. Blue, red, green and black curves correspond to the diameter ratios of 0.3, 0.5 0.6 and 0.9, respectively. Looking at these curves, the logarithm of the activity coefficient decreases very quickly with $x_{1}$, especially for the smaller diameter ratios in which the spheres' diameters are more asymmetric, and for $\sigma_{2}/\sigma_{1} = 0.9$, its feature looks almost a linear (black). The crosses are the SP-MC values taken from Ref.~\cite{Barosova96}. The curves and crosses follow each other.
\begin{figure}[ht]
\centering
    \mbox{\includegraphics[width=0.6000\textwidth]{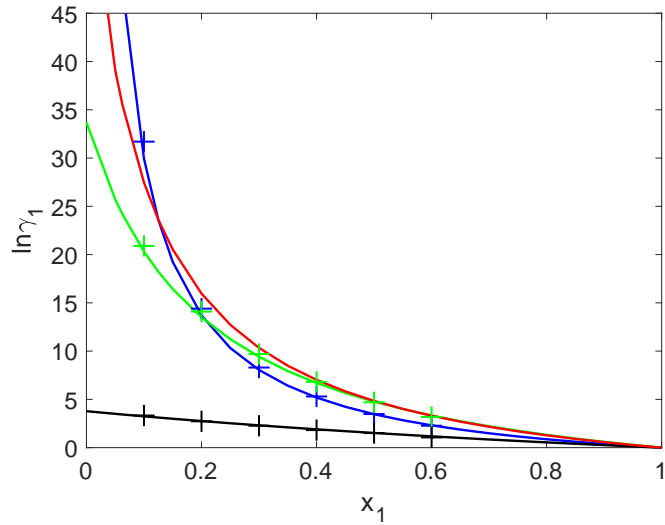}
      }
\caption{ The logarithm of the activity coefficient as a function of the mole fraction $x_{1}$ for the smaller component for $\sigma_{2}/\sigma_{1} = 0.3$ (blue), 0.5 (red), 0.6 (green) and 0.9 (black) and at $\eta = 0.49$. The crosses are results taken from Ref.~\cite{Barosova96}.  } 
\end{figure}
Fig. 2 presents same plots as shown in Fig. 1, but for the larger component. For this case, the logarithm of the activity coefficient is negative. It decreases relatively quickly for smaller diameter ratios as the mole fraction of the larger component increases, which is indeed a similar pattern appeared in Fig. 1. Crosses showing the SP-MC values~\cite{Barosova96} follow our obtained curves.
\begin{figure}[ht]
\centering
    \mbox{\includegraphics[width=0.6000\textwidth]{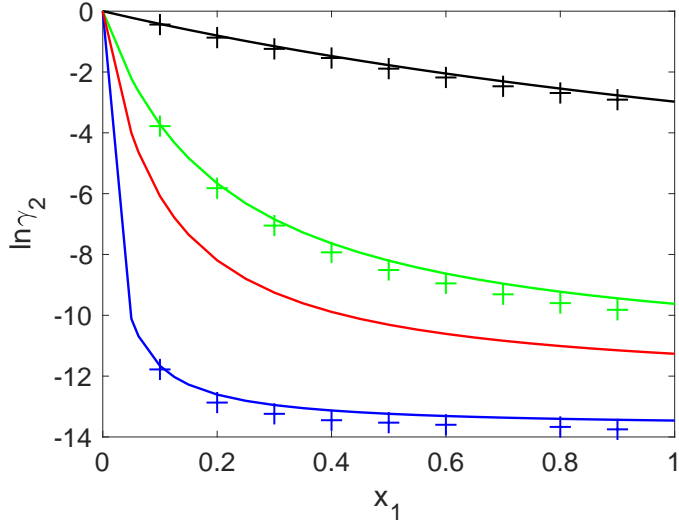}
      }
\caption{ The same plots as shown in Fig. 1, but for the larger component. Crosses are taken from Ref.~\cite{Barosova96}. } 
\end{figure}
Red curves $(\sigma_{2}/\sigma_{1} = 0.5)$ in Fig. 1 and Fig. 2 are not compared with any data. However, looking at other obtained curves, we can note that they can be as accurate as others.

\section{Conclusion}

In this work we have solved the site-site Ornstein-Zernike equation for binary hard-sphere mixture for four different diameter ratios and at $\eta = 0.49$ in the PY approximation. The excess chemical potential for two components as a function of the mole fraction $x_{1}$ of the larger component has been computed with an analytical expression containing the correlation and bridge functions, which in general can be used in an evaluation of the excess chemical potential from the IE approach with any forms of bridge function. Moreover, the logarithms of the activity coefficients for the components as a function of the mole fraction $x_{1}$ have been obtained and compared with accurate SP-MC data. It has been shown that our obtained results agree well with those data.

%\begin{acknowledgments}
%I would like to thank Dr. Danny L. Yeager for helpful discussions. 
%We thank the Robert A. Welch Foundation for financial support from grant A-770.  
%\end{acknowledgments}

\end{document}